 \documentclass[12pt,preprint]{aastex}

\shorttitle{Metallicities of Giants and Subgiants with Planets}
\shortauthors{Ghezzi et al.}

\begin{document}


\title{Metallicities of Planet Hosting Stars: A Sample of Giants and Subgiants\footnote{Based on observations 
made with the 2.2 m telescope at the European Southern Observatory (La Silla, Chile), 
under the agreement ESO-Observat\'orio Nacional/MCT.}}


\author{L. Ghezzi\altaffilmark{1}, K. Cunha\altaffilmark{1,2,3}, S. C. Schuler\altaffilmark{2} \& V. V. Smith\altaffilmark{2}}


\altaffiltext{1}{Observat\'orio Nacional, Rua General Jos\'e Cristino, 77, 20921-400, 
                 S\~ao Crist\'ov\~ao, Rio de Janeiro, RJ, Brazil; luan@on.br}
\altaffiltext{2}{National Optical Astronomy Observatory, 950 North Cherry Avenue, Tucson, AZ 85719, USA}
\altaffiltext{3}{Steward Observatory, University of Arizona, Tucson, AZ 85121, USA}


\begin{abstract}
This work presents a homogeneous derivation of atmospheric parameters
and iron abundances for a sample of giant and subgiant stars which host giant
planets, as well as a control sample of subgiant stars not known to host
giant planets. The analysis is done using the same technique as for our previous 
analysis of a large sample of planet-hosting and control sample dwarf stars.  
A comparison between the distributions of [Fe/H] in planet-hosting main-sequence
stars, subgiants, and giants within these samples finds that the 
main-sequence stars and subgiants have the same mean metallicity of $\langle$[Fe/H]$\rangle \simeq$+0.11 dex, 
while the giant sample is typically more metal poor, having an average 
metallicity of [Fe/H]=$-$0.06 dex. 
The fact that the subgiants have the same average metallicities as the dwarfs indicates that
significant accretion of solid metal-rich material onto the planet-hosting stars has not taken place,
as such material would be diluted in the evolution from dwarf to subgiant.
The lower metallicity found for the planet-hosting giant stars in comparison with the 
planet-hosting dwarfs and subgiants is interpreted as being related to the 
underlying stellar mass, with giants having larger masses and thus, on average
larger-mass protoplanetary disks. In core accretion models of planet formation, 
larger disk masses can contain the critical amount of metals necessary to form giant 
planets even at lower metallicities.

\end{abstract}





\section{Introduction}

\label{int}

A physical property of planetary systems that has yet to be fully understood is
a connection between planetary formation and the metallicities of the host stars.  
There is now unequivocal evidence that main
sequence (MS) FGK-type dwarfs known to have at least one giant planet (i.e.,
M$_{p} \geq 1$ M$_{\mathrm{J}}$, where M$_{p}$ is the planetary mass and
M$_{\mathrm{J}}$ is a Jupiter mass) companion discovered via the radial velocity
method are metal-rich compared to similar stars in the disk field not known to
harbor close-in giant planets \citep[e.g.,][]{s00,s01,s03,s04,fv05,ghezzi10}. 
Contrary to this observation, there is increasing evidence that this planet-metallicity
correlation does not extend to evolved giant stars; giants with planets tend to
be more metal-poor than their main sequence counterparts
\citep[e.g.,][]{schuler05,p07}.

The metallicity distribution of planetary host stars may hold critical clues to
planet formation processes and the subsequent evolution of planetary systems.
Indeed, the favored interpretation of the planet-metallicity correlation
observed for MS dwarfs is that planets form more readily in high-metallicity
environments \citep[e.g.,][]{fv05}, in agreement with predictions
of core accretion planet formation models
\citep[e.g.,][]{il04,ec10}. A competing
interpretation, however, holds that the enhanced metallicity results from the
accretion of H-depleted rocky material onto the star and pollution of the thin
convective envelopes of FGK dwarfs \citep[e.g.,][]{g97}.  This
scenario would be supported by the lower metallicities of giants with planets, which
having been enhanced on the MS, would be diluted by the deepening convection
zones as the stars evolve up the red giant branch.

\defcitealias{ghezzi10}{Paper I}

An observational result that has been used as an argument in favor of the
primordial enrichment hypothesis and against the pollution hypothesis is the
observed metallicities of subgiants: subgiants with planetary companions have
been shown to have enhanced metal abundances, similar to those of MS dwarfs
with planets \citep{fv05}. If the difference in metallicities of
planet hosting dwarfs and giants results from the pollution and subsequent
dilution of the stars' convective envelopes, one might expect the subgiants to
have intermediate metallicities, forming a metallicity gradient from the
metal-rich MS dwarfs, to the increasingly diluted subgiants, and finally to the
fully diluted giants.  Heretofore, this pattern has not been observed.  In this
paper, we present the results of a homogeneous metallicity ([Fe/H])
analysis of 15 subgiants and 16 giants with planetary companions, as well as a
control sample of 14 subgiants not known to harbor closely orbiting giant planets.
This sample of evolved stars, which includes both giants and subgiants, constitutes the first 
to be analyzed in a homogeneous fashion within a single study.  These metallicities
are compared to those of a large sample of main sequence dwarfs with and without 
planets that have been derived as part of the same analysis and have been recently reported
in \citet{ghezzi10} \citepalias{ghezzi10}.


\section{Observations}

\label{sample}

The sample of planet hosting stars studied here contains 31 targets. 
The target list was compiled from the Extrasolar Planet Encyclopaedia\footnote{Available at http://exoplanet.eu},
and these stars were originally part of the larger sample analyzed in \citet{ghezzi10}: 
the latter study focused on the analysis of dwarf stars while the more evolved objects, giants and
subgiants, are presented here.
A sample of disk subgiants (N=14) observed to not host closely orbiting giant planets was 
also observed with the same set-up, and the target list was obtained from the list of candidates
deemed to be \textquotedblleft RV stable\textquotedblright\ from \citet{fv05}. The list with all stars analyzed in this study can be found in Table \ref{obslog}. 

The observations consist of high-resolution spectra (R $= \lambda/\Delta \lambda \sim$ 48,000) 
obtained with the FEROS spectrograph (\citealt{kaufer99}) MPG/ESO-2.20 m telescope (La Silla, 
Chile)\footnote{Under the agreement ESO-Observat\'orio Nacional/MCT.}. 
The spectra were reduced in a standard way. A more detailed account of the observations and
the data reduction is provided in \citetalias{ghezzi10}. 
A log of the observations with $V$ magnitudes, observation dates, integration times and 
signal-to-noise ratios can be found in Table \ref{obslog}. 

\section{Stellar Parameters and Metallicities}

\label{param_feh}

The derivation of stellar parameters and metallicities ([Fe/H]) in this study followed the
same methodology presented and discussed in \citetalias{ghezzi10}. 
The same selection of \ion{Fe}{1} and \ion{Fe}{2} lines was analyzed and their equivalent widths
were also measured using the automatic code of equivalent width measurement ARES (\citealt{s07}). 
In order to further test the quality of automatic equivalent width measurements
for the parameter space covered by this particular set of subgiant and giant stars, 
equivalent widths of two sample targets 
HD 188310 (with T$_{eff}$ typical of the giants in our sample and a spectrum with high S/N) and  HD 27442 
(typical T$_{eff}$ but with a lower S/N spectrum) were measured manually (using the task \texttt{splot} on IRAF). 
Our results indicate that equivalent widths measured with IRAF compare favorably with the automatic ones: 
$\langle$EW$_{ARES}-$EW$_{Manual}\rangle$ =  $-0.43 \pm 2.28$ m\AA\ for HD 188310
and $+0.15 \pm 3.42$ m\AA\ for HD 27442, which is consistent with previous results in \citet{s07}.

Effective temperatures, surface gravities, microturbulent velocities and iron abundances were derived 
under the assumption of LTE and self-consistently from the requirement that
the iron abundance be independant of the line excitation potential and measured equivalent widths,
as well as from the forced agreement between \ion{Fe}{1} and \ion{Fe}{2} abundances. 
Although this analysis uses the approximation of LTE, a discussion of possible non-LTE effects
in the Fe abundances will be presented in Sextion 4.1.3.
Table \ref{atm_par} lists the derived stellar parameters for the target stars. 
The number of \ion{Fe}{1} and \ion{Fe}{2} lines (and the standard deviations in each case) for each star is also listed.

Uncertainties in the derived parameters $T_{\rm eff}$, log g, $\xi$ and [Fe/H] can be estimated as in
\citet{gv98}, similarly to \citetalias{ghezzi10}. 
The typical values for the internal errors in this study are $\sim$ 50 K in $T_{\rm eff}$, 
0.15 dex in log g, 0.05 km s$^{-1}$ for $\xi$, and 0.05 dex in [Fe/H]. (See \citetalias{ghezzi10} for a
discussion of these internal uncertainties). We note, however, that the real uncertainties are expected to
be somewhat larger (100 K in $T_{\rm eff}$, 0.20 dex in log g, 0.20 km s$^{-1}$ in $\xi$ and 0.10 dex in [Fe/H])
than the internal errors. 
The sensisitvity of \ion{Fe}{1} abundances to changes in the parameters $T_{\rm eff}$, log g and $\xi$
is also investigated. 
For this exercise, we use 2 giants that span the $T_{\rm eff}$ interval of most of the giant sample: 
HD 11977 ($T_{\rm eff}$ = 4972 K), NGC 2423 3 ($T_{\rm eff}$ = 4680 K), and the subgiant HD 11964 ($T_{\rm eff}$ = 5318 K). 
A variation of $\pm$100 K in $T_{\rm eff}$ induces a change of $\pm$0.03 and $\pm$0.05 dex in A(\ion{Fe}{1}) 
for the coolest (HD 122430) and hottest (HD 11977) giants, respectively. For the subgiant, the sensitivity is $\pm$ 0.09 dex.
A variation of $\pm$0.2 dex in log g does not affect significantly the Fe abundances: A(Fe) changes by  $\sim$0.01 dex
for the hotter stars and $\sim$0.03 dex for the cooler stars.
As expected, a decrease in the microturbulence causes an increase in A(\ion{Fe}{1});
for the subgiant star, a change of $\pm$0.20 km s$^{-1}$ causes a variation of $\mp$0.08 dex in the \ion{Fe}{1} abundance,
while for the giants, this variation is around $\mp$0.10.
The total errors in A(Fe) from these typical uncertainties are $\pm$ 0.11 dex for the subgiants and $\pm$ 0.13 dex for the giants.
As the results in this study for the giants and subgiants will be compared to those 
for the dwarfs in \citetalias{ghezzi10}, we repeat the above exercise for 
a typical dwarf with solar parameters (HD 106252; \citetalias{ghezzi10}). 
Variations of $\pm$100 K in $T_{\rm eff}$, $\pm$0.20 dex in log g and $\pm$0.20 km s$^{-1}$ in $\xi$ cause changes of, 
respectively, $\pm$0.08, $\leq$0.01 and $\mp$0.05 dex in A(\ion{Fe}{1}); or a total error of $\pm$ 0.1 dex. 
These total uncertainties for the dwarfs in Paper I are slightly lower but not significantly different from the 
total uncertainties estimated for the subgiants (0.11 dex) and giants (0.13 dex). 

\begin{figure}
\epsscale{0.70}
\plotone{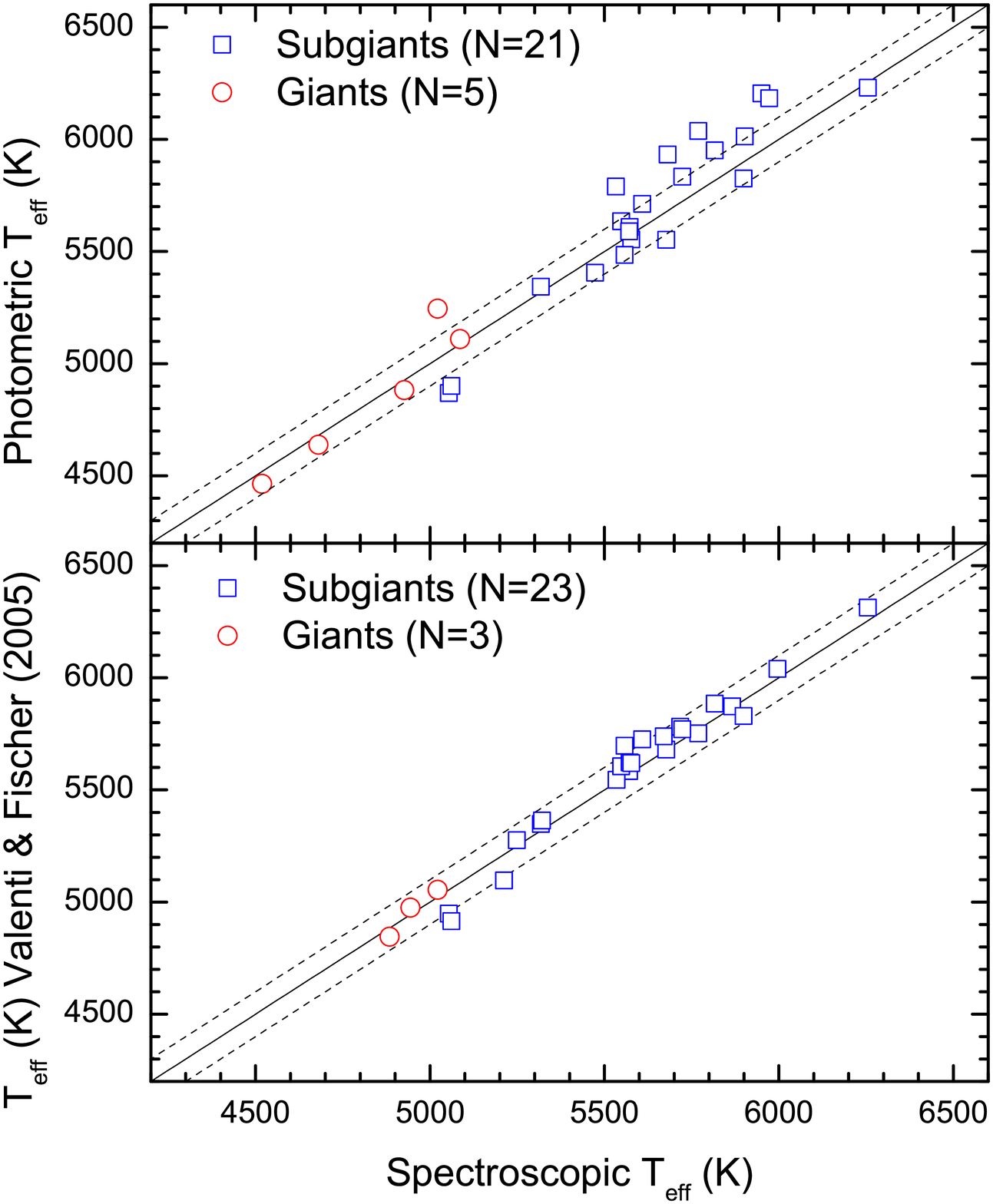}
\caption{Top Panel: Comparison between the spectroscopic effective temperatures derived in this study with
$T_{eff}$'s derived from the $V-K$ calibration by \citet{ghb09}. Subgiants are the open blue squares and 
giants are the open red circles. Bottom Panel: Comparison between the effective temperatures derived in 
this study with the stars in common with \citet{vf05}. 
The solid line represents perfect agreement and the dashed lines $\pm$100 K. 
The three effective temperature scales shown in the top and bottom panels show good agreement 
within the expected errors in the determinations.}
\label{tefs}
\end{figure}

The derived effective temperatures for the stars in our sample can be compared with 
independent results from photometric $V-K$ calibrations. 
Several photometric calibrations are available in the literature (e.g. \citealt{a99}; \citealt{rm05}). 
\citet{ghb09} presented a new implementation of the infrared flux method using 2MASS magnitudes and Kurucz models. 
A comparison of the derived spectroscopic effective temperatures with their photometric calibration is 
shown in the top panel of Figure \ref{tefs}. Results are shown for all stars in our sample which 
have unsaturated $K_{s}$ 2MASS magnitudes (with errors in $K_{s} <$ 0.1 mag); reddening corrections 
from \citet{a92} were applied to obtain the de-reddened colors. 
The comparison between the two scales is quite good for the entire $T_{\rm eff}$ range, with agreement 
for most of the stars within $\pm$100 K (shown as the dashed lines in the figure). There is not a significant 
systematic difference in the effective temperatures between giants and subgiants, but we note a few 
outliers falling above the dashed lines (mostly subgiants) and 2 results for  subgiants which fall below. 
The average difference between the two scales is well within the expected errors and overall agree 
with the variations typically found between different $T_{\rm eff}$ scales in the literature:
$\langle\delta T_{eff}$ (This Study - \citealt{ghb09})$\rangle$= $-48 \pm 136$ K.

The more recent calibration by \citet{c10} is hotter than that of \citet{ghb09}; 
a comparison with our results for the subgiants (the calibration of \citealt{c10} only applies for dwarfs and subgiants) 
shows a larger systematic difference of 
$\langle\delta T_{eff}$ (This Study - \citealt{c10})$\rangle$= $-106 \pm 147$ K. 
For the calibration of \citealt{ghb09}, we find $\langle\delta T_{eff}\rangle_{subgiants} = -55 \pm 142$ K. 
This difference of $\sim$50 K is consistent with the discussion presented in section 2.5 of \citet{c10}. 
Reddening corrections applied to (V-K) influence the derived photometric temperatures: 
a change of 0.01 mag in E(B-V) can lead to a change of 50 K in the effective temperature (\citealt{c10}). 
Note that \citet{c10} have not adopted reddening corrections for stars in their samples closer than $\sim$75 pc. 
If we also neglect reddening  corrections for those stars in our sample which are closer than $\sim$75 pc and 
recompute the photometric $T_{eff}$s, the average differences become 
$\langle\delta T_{eff}$ (This Study - \citealt{c10})$\rangle$= $-33 \pm 144$ K.


The bottom panel of Figure \ref{tefs} shows the comparison of our derived effective temperatures 
with those obtained by \citet{vf05}; the latter study also derived $T_{eff}$ spectroscopically, 
although their analysis followed a different method which consisted in fitting the observed spectra 
by adjusting 41 free parameters (one of them being the effective temperature). 
The $T_{eff}$ results in the two spectroscopic analyses agree well: 
$\langle\delta T_{eff}$ (This Study - \citealt{vf05})$\rangle$= $-18 \pm 67$ K.

\subsection{Evolutionary Parameters}

\label{evol}

\begin{figure}
\epsscale{1.00}
\plotone{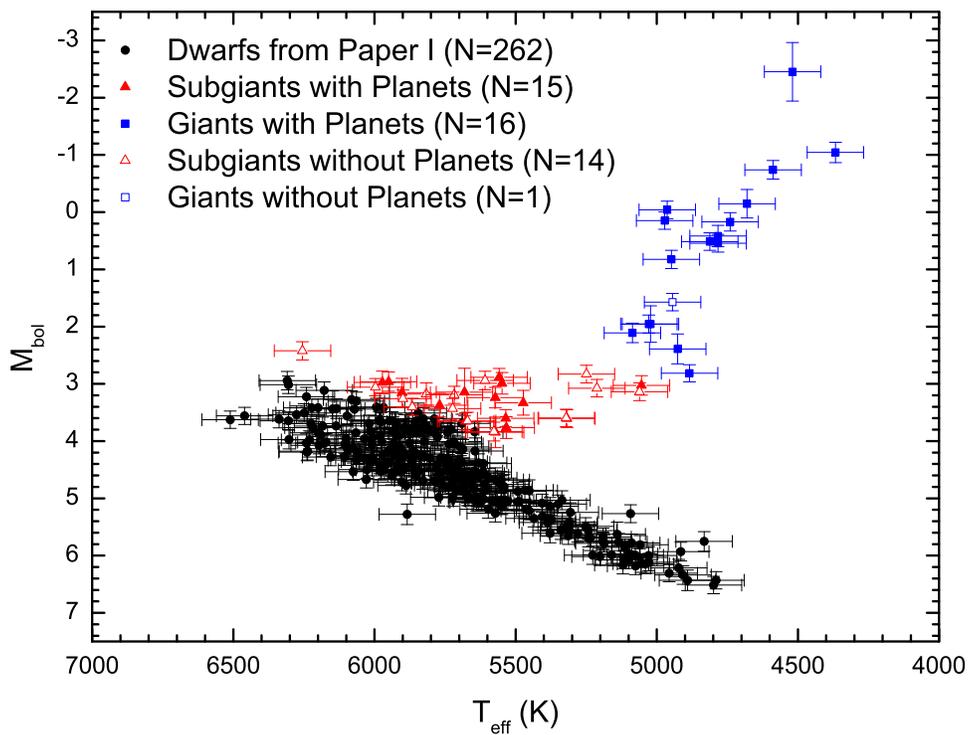}
\caption{Location of studied stars in an H-R diagram. 
The targets analysed in this study are evolved away from the main sequence. 
The samples are segregated in dwarfs (black circles; analyzed in \citetalias{ghezzi10}), subgiants (red triangles) and giants (blue squares).
All giant stars in our sample, except one, host giant planets; the sample of subgiants include both 
planet hosting stars as well as a control sample of subgiant stars known to not host giant planets.}
\label{Mbol_teff}
\end{figure}

As mentioned previously, \citetalias{ghezzi10} analyzed unevolved stars with and without planets,
while the present study focuses on more evolved stars, also both with and without giant planets.
Figure \ref{Mbol_teff} shows the location of the sample stars in an HR diagram with the bolometric magnitudes versus effective temperatures. 
The bolometric magnitudes for the stars were calculated using the bolometric corrections of \citet[see details 
in \citetalias{ghezzi10}]{g02}. This figure also includes for comparison the sample of stars which was studied in \citetalias{ghezzi10} (represented 
by black filled circles), and these generally define the location of the main sequence. 
The target stars analyzed here are obviously more evolved. 
In this study, a star is classified as a subgiant (represented as red triangles in Figure \ref{Mbol_teff}) if it is 1.5 mag above the lower boundary of the main sequence and 
has $M_{bol} > 2.82$ ; the 17 stars which have $M_{bol} < 2.82$ are classified 
as giants (represented as blue squares in Figure \ref{Mbol_teff}). This boundary transition between the main-sequence and
the subgiant branch is somewhat uncertain and for two stars in particular we adopted a different classification:
HD 2151 was classified as a subgiant (although it is not 1.5 mag above the lower boundary of the main sequence) because 
of its low derived values of log g ($\sim$ 4.0), and HD 205420 (the isolated star with $T_{eff}$= 6255 K and $M_{bol} < 2.82$) is 
considered as a subgiant. The transition between the subgiant and giant branches is also uncertain. In particular,
the classification of two stars (HD 177380 and HD 208801) which lie close to the base of the red-giant branch is uncertain; however, their derived
surface gravities are more compatible with their classification as subgiants. (Nevertheless, the implications of including these stars as
giants in our analysis is discussed in Section \ref{sg}). The adopted classification of the sample stars in giants and subgiants can be
found in the last column of Table \ref{obslog}. 


Table \ref{evol_par} summarizes the evolutionary parameters calculated for the studied stars. 
The parallaxes are from the Hipparcos catalogue; the luminosities are calculated using the parallax,
V magnitudes, reddening and the derived effective temperatures (see \citetalias{ghezzi10} for details).
Three stars (namely NGC 2423 3, NGC 4349 127 and HD 171028) were not present in the Hipparcos catalogue, thus 
their $V$ magnitudes and parallaxes come from the references in the Extrasolar Planet Encyclopaedia. Note also 
that the \citet{a92} model for reddening is accurate to distances within 1 kpc of the Sun.
The radii, masses, as well as Hipparcos gravities and estimated ages in Table \ref{evol_par} were calculated using L. Girardi's
web code PARAM\footnote{Available at http://stev.oapd.inaf.it/cgi-bin/param}, which is based on a Bayesian parameter estimation method (\citealt{dS06}). We note
that the $Y^{2}$ evolutionary tracks (\citealt{y03}) were not used in this paper because these do not follow evolution through the red clump.

\begin{figure}
\epsscale{1.00}
\plotone{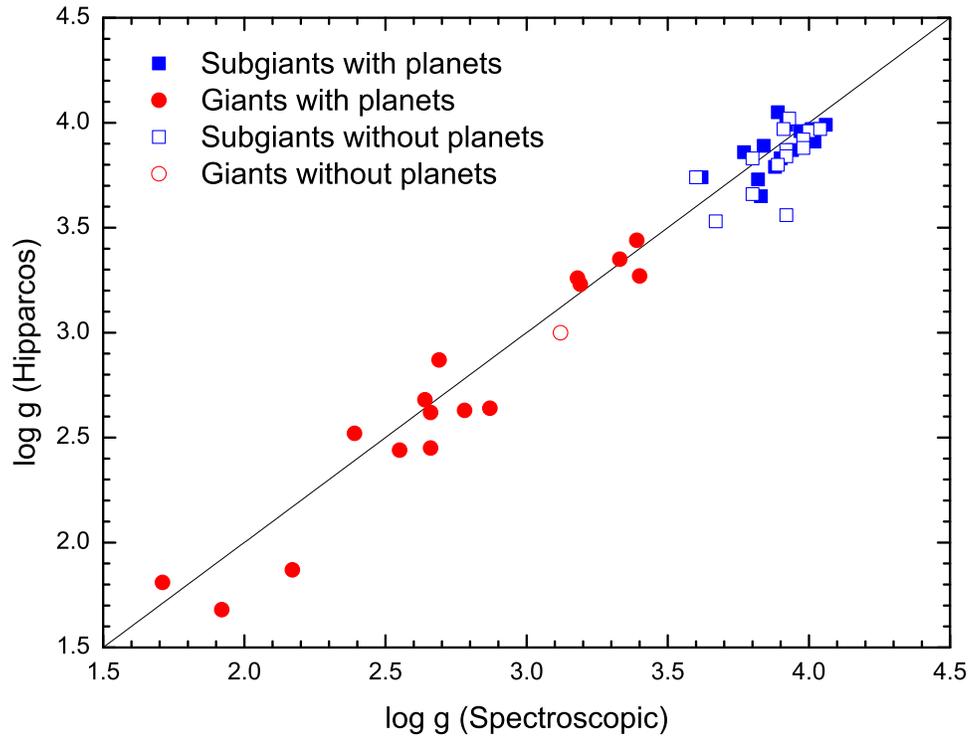}
\caption{Comparison between the spectroscopic gravities derived in this study with those derived using
Hipparcos parallaxes. The agreement between the two scales is good with no significant offsets. Perfect
agreement is represented by the solid line.}
\label{loggs}
\end{figure}

The surface gravity values obtained here from the ionization equilibrium of \ion{Fe}{1} and \ion{Fe}{2} 
and in LTE (column 3; Table \ref{atm_par}) can be compared with gravities which are based on Hipparcos parallaxes (column 11; Table \ref{evol_par}); 
such a comparison is shown in Figure \ref{loggs}. The agreement between the average values for the two scales 
is found to be good: $\langle\delta$ (log g Hipparcos $-$ log g This Study)$\rangle$ = $-0.04 \pm$ 0.12 dex.  Such an agreement between the Hipparcos gravities and spectroscopic values derived from the agreement between Fe I and Fe II suggest the absence of strong non-LTE effects (see discussion in Section \ref{nlte}).


Concerning their masses and ages, the sample of giants studied here is more massive and younger than the subgiants
and dwarfs (from \citetalias{ghezzi10}). The giants in our sample have an average mass of 1.82 $\pm$ 0.68 M$_{\sun}$ and a
distribution ranging between $\sim$ 1.1 -- 3.8 M$_{\sun}$; their average age is $\langle Age\rangle_{giants}$= 2.22 $\pm$ 1.37 Gyr. 
For comparison we note that the sample dwarfs (\citetalias{ghezzi10}) have $\langle M\rangle_{dwarfs}$= 1.03 $\pm$ 0.17 M$_{\sun}$ (encompassing the interval $\sim$ 0.6 -- 1.4 M$_{\sun}$) and $\langle Age\rangle_{dwarfs}$= 5.34 $\pm$ 2.70 Gyr. The overlap in the mass range between the samples of giants and dwarfs is therefore small. We note that the masses and ages of the dwarfs were derived in a different way (Y$^{2}$ evolutionary tracks and isochrones). However, it was shown in \citetalias{ghezzi10} that the results from the two methods are consistent for dwarfs: $\Delta $M (Y$^{2}$ - Girardi's Code) = 0.03 $\pm$ 0.05 $M_{\sun}$ and $\Delta$ Age (Y$^{2}$ - Girardi's Code) = 0.37 $\pm$ 1.46 Gyr (see last paragraph of Section 3.4 in \citetalias{ghezzi10}).

In terms of their average masses, the sample of subgiants studied here falls technically in between the sample of giants and dwarfs but 
there is considerable overlap in the mass range of the dwarf sample (the subgiant sample encompasses the interval $\sim$ 1.0 -- 1.5 M$_{\sun}$; with an average mass of 1.20 $\pm$ 0.14 M$_{\sun}$).
It is interesting to note that the subgiants in our sample are on average slightly older than the dwarfs ($\langle Age \rangle_{subgiants}$ = 5.46 $\pm$ 1.92 Gyr),
representing the oldest population in this study. In summary, the sample subgiants are a more evolved
population of the previously studied dwarfs from \citetalias{ghezzi10} as these dwarfs and subgiants have approximately the same
mass ranges. The sample giants, however, are evolved from stars which are more massive and
are on average the youngest of all target stars. 


\section{Discussion}

\label{disc}

\subsection{Metallicity Distributions of Evolved Stars Hosting Planets}

As the number of discovered planet hosting stars increases and samples include a larger number of stars which
are on the red-giant branch, metallicity distributions of evolved stars hosting planets have started to appear 
in the literature. Because the samples are still relatively small, and the abundance analyses are not always
homogeneous, there is some controversy in some of the conclusions of recent studies of giants,
which are briefly summarized as follows. \citet{schuler05} compared the iron abundances of
7 giants with planets known at the time and found that their metallicity distribution was on average
lower than that of dwarfs with planets. This result was later confirmed by 
\citet{p07} who concluded that the metallicity distributions of giants with planets 
do not favor metal-rich systems. Their results were based on a sample of 14 giants with planets 
(10 of which analyzed by their group) and are interpreted as possible evidence for pollution. 
\citet{hm07} analyzed 380 G-K giants as part of the radial velocity survey at Lick Observatory. 
Five of these stars host planets; they also gather abundances from the literature for another 15 giants and 
obtain an average metallicity for the sample of -0.05 dex. In addition, this study concludes that
there is an offset of 0.13 dex between the metallicity distributions of giants with and without planets;
the latter are found to be generally more metal poor.
Such an offset in the metallicity distributions is not confirmed in the recent study by \citet{t08}, who 
analyzed a sample of 322 intermediate-mass late-G giants; ten of these stars host planets. Their 
comparisons between the metallicity distributions of giants with and without detected planets reveals no 
significant difference between the two samples; both distributions have average metallicities around 
-0.12 dex. In the following sections we discuss the metallicity distributions for the giant and subgiant
planet-hosting stars in our sample.

\subsubsection{Giants}

\label{g}

\begin{figure}
\epsscale{0.80}
\plotone{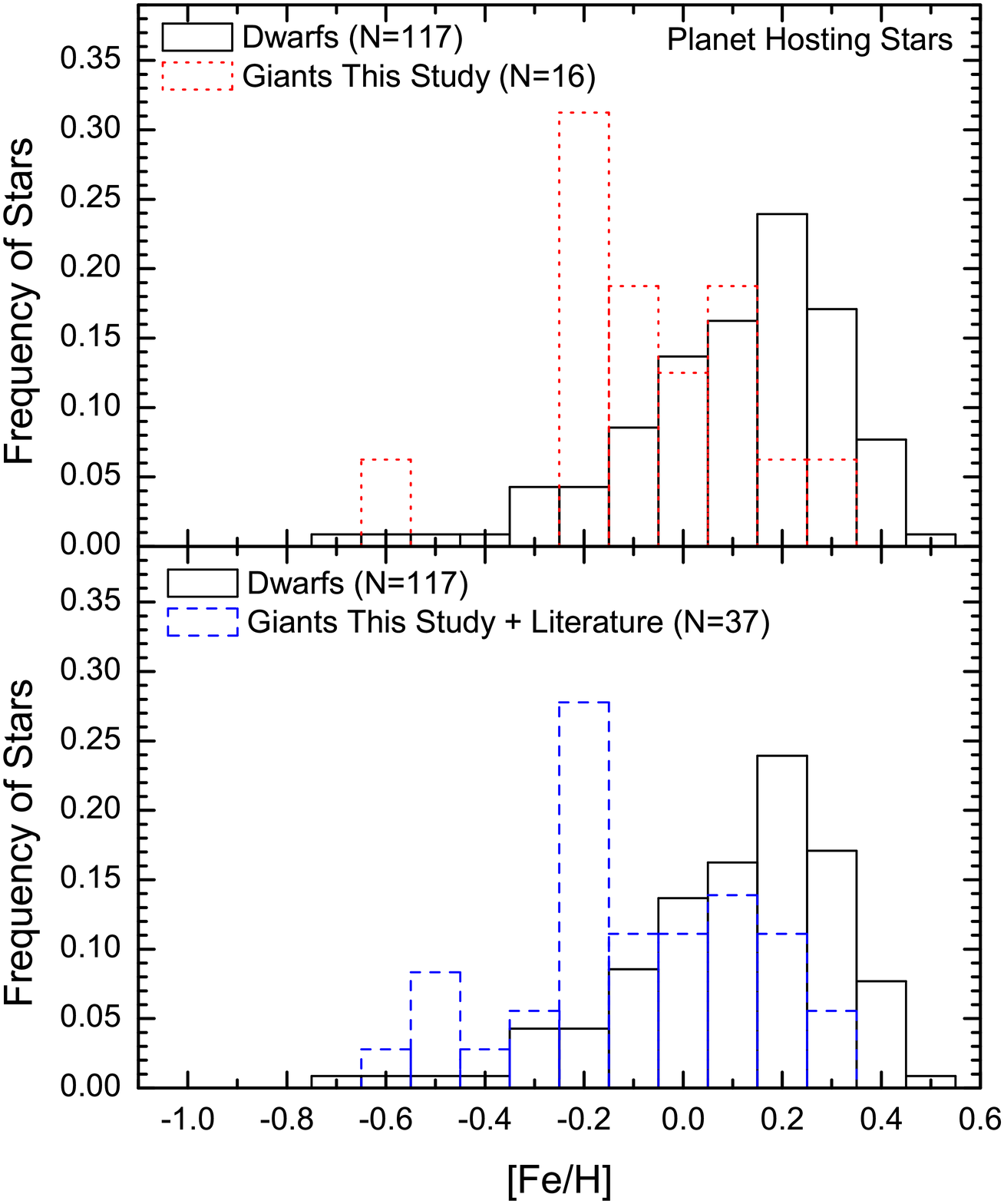}
\caption{Top panel: Metallicity distributions obtained for planet hosting dwarfs (black solid line), and giants (red dotted line). All
abundance results in these distributions were derived homogeneously. Bottom panel: Metallicity distributions for planet hosting dwarfs (black solid line; same as top panel), and all giant star hosting giant planets known
to date (blue dashed line). The metallicities for those planet-hosting giants not analyzed in this study are taken
as the average of the iron abundance values found in the literature (see Table \ref{giant_lit}).}
\label{giant_hist}
\end{figure}

The iron abundance distribution derived from the sample of giant stars hosting planets (N=16) is shown 
in the top panel of Figure \ref{giant_hist} as a red dotted line histogram; the average value for this distribution is $\langle$[Fe/H]$\rangle_{giants}$= -0.06 dex. 
For comparison, the iron abundances obtained for the sample of dwarf stars hosting giant planets (N=117) from \citetalias{ghezzi10} are also shown
(black solid line histogram). It is apparent from the figure that the metallicity distribution of the giant stars peaks at a lower 
metallicity value when compared to the dwarf stars; the difference between the average [Fe/H] is 0.17 dex.
The application of a Kolmogorov-Smirnov (K-S) test gives a probability of only 1\% that the main-sequence dwarfs and giants are 
drawn from the same parent Fe-abundance population.
It is important to recognize, however, the relatively small number of giant stars in this comparison, although this contains 
$\sim$45\% of the total number of giant stars hosting giant planets found to date.


In order to improve the giant star statistics as much as possible, iron abundances for the remaining giants known to have planets
were collected from different studies in the literature; their metallicities are listed in Table \ref{giant_lit}. Literature results for the giant star sample studied here are also presented for comparison.
The histogram in the bottom panel of Figure \ref{giant_hist} shows the metallicity distribution for the sample including both the giants in this study and for all other literature giants in Table \ref{giant_lit} (N=37; blue dashed line). The metallicity distribution for the planet-hosting dwarfs from \citetalias{ghezzi10} is shown again for comparison. Using the extended giant sample
yields a similar conclusion: the metallicities of giant stars with planets are on average lower than dwarfs with planets.
In particular, the average metallicity for the extended sample (giants from this study plus literature) is somewhat 
 lower ($\langle$[Fe/H]$\rangle$=-0.12 dex),  but not significantly so, 
 than the average metallicity obtained using only the giants from this study by 0.06 dex.
The results from this extended giant sample reinforce the premise that giant stars with giant planets seem to have on average lower 
iron abundances than dwarf stars with giant planets. The application of a K-S test using the extended giant sample gives a very
small probability of 1.91 x 10$^{-5}$\% that the main-sequence dwarfs and giants are drawn from the same parent population. 
Such results are in line with the conclusions by \citet[see also \citealt{schuler05}]{p07} who discuss that 
the metallicity distributions of planet hosting dwarfs and giants are different; with the giant stars 
having a distribution shifted to lower metallicites by 0.2--0.3 dex with respect to the dwarfs. 

As a final note, we recall the discussion about uncertainties in Section \ref{param_feh}. It was shown that our spectroscopic temperatures for the giants would be $\sim$20 K cooler if we considered the photometric temperatures as the \textquotedblleft correct\textquotedblright\ scale. This would result in underestimated \ion{Fe}{1} abundances by 0.02 dex at most. The sensitivity of this parameter to $T_{\rm eff}$, log g and $\xi$ was also discussed and the conclusion was that relative systematic effects of up to 0.1 dex can exist when comparing abundances of dwarfs and giants. Comparisons of our metallicities with those from many studies in the literature (see Table 5 of \citetalias{ghezzi10} and Table \ref{giant_lit} of this study) do not show evidence for these possible systematic effects in our metallicities. Even if they existed, neither would be sufficient to explain the differences of 0.17 -- 0.23 dex found between the average metallicities of dwarfs and giants. 

\subsubsection{Subgiants}

\label{sg}

\begin{figure}
\epsscale{0.70}
\plotone{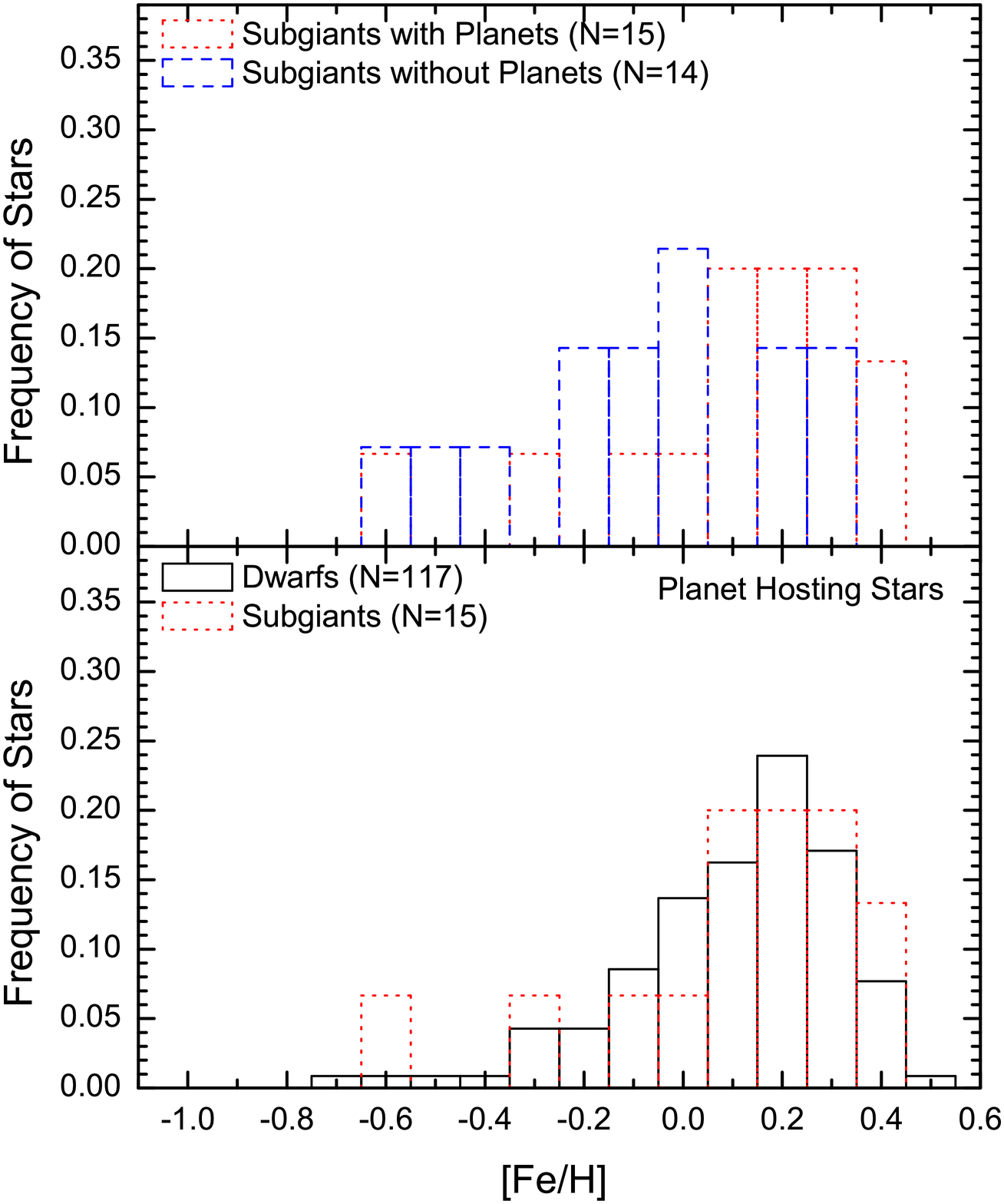}
\caption{Top panel: Metallicity distributions obtained for the sample of subgiant stars hosting planets (red dotted line histogram) and the control sample of subgiants not known to have giant planets (blue dashed line histogram). 
The planet-hosting stars
are found to be on average more metal rich than the control sample by 0.21 dex. Bottom panel: Metallicity distributions
of subgiant stars with planets (red dotted line histogram) in comparison with the dwarf star planet hosting sample (black solid line histogram) analyzed in \citetalias{ghezzi10}. The two
distributions are similar with no obvious abundance shifts.}
\label{subgiant_hist}
\end{figure}

An additional important aspect of the present study is the homogeneous abundance analysis for samples of subgiants 
with and without giant planets. A comparison of the metallicity distributions for the two samples
(Figure \ref{subgiant_hist} top panel) points to a similarity to that found for dwarfs; namely 
that subgiant stars without planets are on average more metal poor than the sample of subgiants hosting planets. 
The results from \citetalias{ghezzi10} showed that the metallicity distribution of dwarfs hosting planets was
more metal rich by 0.15 dex than that for dwarfs not hosting planets. This abundance offset found previously for 
unevolved stars compares well with the difference obtained here for subgiants: the 
average metallicity for our sample of subgiants hosting planets (N=15) is $\langle$[Fe/H]$\rangle$ 
= +0.12 dex and for subgiants without planets (N=14) it is more metal poor by 0.21 dex 
($\langle$[Fe/H]$\rangle$ = -0.09 dex).
\citet{fv05} also analyzed a sample of 86 subgiants, nine of which host giant planets.
They find that the median metallicity of their sample of subgiants without detected planets is -0.01 dex, while
that of the sample of subgiants with planets is +0.35 dex. This is more metal rich than the results
found here for planet hosting subgiants: the median metallicty of the subgiant distribution obtained here is 0.20 dex.


As discussed in the previous section, the sample of subgiants studied here is in fact on average older than
the sample of dwarfs (from \citetalias{ghezzi10}), as well as the giant star sample. In terms of their mass distribution the subgiants,
although including a few more massive stars, constitute the same general population as the dwarfs,
but are just older and more evolved. 
A comparison of the metallicities of subgiants in our sample and dwarfs from \citetalias{ghezzi10}
is shown in Figure \ref{subgiant_hist} (bottom panel), and the distributions are not
significantly different. A K-S test gives a probability of 56\% that the
two samples belong to the same parent population. 
Based on a K-S test applied to their samples, \citet{fv05} also find that the 
metallicity distributions of main-sequence and subgiant stars with planets are consistent, and that 
both samples are more metal-rich than their counterparts without detected planets.

There is presently a negligible offset (0.01 dex) between the averages of the metallicity distributions of dwarfs with planets (from \citetalias{ghezzi10})
and the subgiants with planets studied here: both have $\langle$[Fe/H]$\rangle \simeq$+0.11 dex. In general terms, this is what
would be expected if dwarfs and subgiants come from the same population if there are no effects related to
age-metallicity. It should be recognized, however, that the 
subgiant sample is significantly smaller than the dwarf sample and that this offset in metallicity, which
is found to be zero for the stars with planets, could in fact be as large as $\sim$0.05 dex given the uncetainties 
in the analysis and the small number statistics. In fact, there is a small offset of 0.05 dex between the 
averages of the metallicity distributions of dwarfs and subgiants without planets (with the latter being more metal poor). For example,
if the the planet hosting star (HD 177830) which lies in the transition between the subgiant and red-giant branches
(previously noted in Section \ref{evol}; Figure \ref{Mbol_teff}) is instead classified as a giant (as in \citealt{hm07}), 
this will affect  the average metallicity of the sample of subgiants with planets which will change to
a slightly lower value: $\langle$[Fe/H]$\rangle$=+0.10 dex (N=14); or an offset between subgiants and dwarfs with planets
of 0.01 dex (note also that in this case the giants will have an average metallicity which is slightly higher of  $\langle$[Fe/H]$\rangle$=-0.03 dex).
If this offset in the metallicities between the dwarfs and subgiants is small but real, a possible interpretation 
for the lower metallicity found for the subgiants is that these small differences in the abundances are the result of 
chemical evolution, since the sample subgiant stars are older they would be on average slightly more metal poor. 

We note that the discusson about uncertainties presented in Section \ref{param_feh} revealed that no significant systematic effects should be expected in the comparison of metallicities of subgiants and dwarfs. It was also shown that our spectroscopic temperatures for the subgiants would be $\sim$50 K cooler if we considered the photometric temperatures as the \textquotedblleft correct\textquotedblright\ scale. This would result in underestimated \ion{Fe}{1} abundances by 0.05 dex at most. Therefore, if possible systematic effects do exist in our metallicities (which does not seem to be true), they would not change the main point of the above discussion: the average metallicities of dwarfs and subgiants are equal within the expected uncertainties.

\subsubsection{Departures from LTE in \ion{Fe}{1} and \ion{Fe}{2} in Dwarfs, Subgiants, and Giants}

\label{nlte}

Given the comparisons in the metallicity distributions
(primarily from \ion{Fe}{1} but also from \ion{Fe}{2} lines) 
between the dwarf, subgiant, and giant samples discussed here, 
it is important to assess non-LTE \ion{Fe}{1} and \ion{Fe}{2}
line-formation as a function of T$_{\rm eff}$, surface gravity, and
stellar metallicity.  Within the homogeneous LTE analysis conducted in
this study, planet-hosting dwarf and subgiant stars display the
same [Fe/H] distributions, while there is an overall difference of 
$\sim$ 0.2 dex in
the Fe-abundance distributions of dwarfs and subgiants with planets
when compared to giants with planets. Could this difference be due
simply to different non-LTE effects between dwarfs/subgiants and giants?

\subsubsubsection{Non-LTE Calculations}

Non-LTE calculations for iron contain 
uncertainties due to such quantities as
electronic collisional cross-sections, in particular
for dipole-forbidden transitions; photoionization
cross-sections, in particular those from the excited states;
a treatment of upper states (particularly those for which
no laboratory-measured energies are available); recombination
to the upper levels; a treatment of autoionizing levels
and related photoionization resonances; and, most importantly, 
uncertainties due to treatment of collisions with neutral hydrogen atoms,
which are poorly known (Hubeny 2010, private communication). 
Within these uncertainties, however, results
from non-LTE calculations in cool stars (e.g., \citealt{m10a}; \citealt{g01a,g01b}) 
generally find that departures from LTE become larger in
very metal poor stars, evolved stars, and stars with effective temperatures T$_{eff}>$ 6000 K. 
In the temperature and gravity regimes considered here, non-LTE departures are much less important 
for \ion{Fe}{2} lines, which are generally found to be closer to LTE.

Certain studies of non-LTE in \ion{Fe}{1} and \ion{Fe}{2} find rather
small departures from LTE, even in rather metal-poor stars,
such as globular cluster stars.  For example, \citet{k03}
analyze main-sequence turn-off stars, subgiants, and
giants in the globular cluster NGC 6397 (with [Fe/H]= -2.35)
and find total corrections to LTE of \ion{Fe}{1} of only 0.03 -
0.05 dex, with no differential non-LTE effects between the
main-sequence turn-off and giant stars.  A more recent
analysis of this cluster by \citet{lind09} notes that
such small corrections found by \citet{k03} were 
probably due to their adoption of rather high efficiencies
of collisions between iron atoms and neutral hydrogen atoms,
as paramaterized by the H I collision enhancement factor of
S$_{\rm H}$ = 3. 

Given that one of the major uncertainties in non-LTE calculations, as discussed above,
is the efficiency of collisions of \ion{Fe}{1} with neutral hydrogen,
\citet{m10a} present results for H I collision enhancement factors, S$_{\rm H}$, 
varying between 0 (which corresponds to the strongest non-LTE case) and 2 (corresponding to
a situation closer to LTE), as well as LTE.  Figure 1 in their study  
illustrates differences between non-LTE \ion{Fe}{1} and \ion{Fe}{2} abundances for 4 different stars: Procyon 
(T$_{\rm eff}$ = 6510 K, log g = 3.96, [Fe/H] = -0.10), $\beta$ Vir (T$_{\rm eff}$ = 
6060 K, log g = 4.11, [Fe/H] = +0.04), $\tau$ Cet (T$_{\rm eff}$ = 5377 K,
log g = 4.53, [Fe/H] = -0.43), and HD 84937 (T$_{\rm eff}$ = 6350 K, log g = 4.00,
[Fe/H] = -1.94).  It is clear from this figure that non-LTE \ion{Fe}{1} abundances
of the most metal poor stars in the sample can be affected by as much as 0.15 dex when S$_{\rm H}$ = 0, 
however this is
only for the most metal-poor star in the sample, HD 84937; the effect of
non-LTE on \ion{Fe}{1} decreases significantly for increasing metallicities, due to increasing electron densities.  
In addition, values of S$_{\rm H}$ as low as 0.1 lead to very small non-LTE
corrections for \ion{Fe}{1} for all stars. Although uncertain, the value S$_{\rm H}$=0.1 is 
favored (\citealt{m10c}) 
and this would suggest that differences in the iron abundances between the
samples of near-solar metallcity dwarfs, subgiants, and giants caused by
non-LTE corrections would be less than 0.1 dex.

The calculations presented in \citet{m10a}
predict that non-LTE corrections for \ion{Fe}{1} increase strongly with
decreasing metallicity and, therefore, should be minimal at solar metallicities.  
In addition, the corrections become increasingly
important for effective temperatures greater than T$_{\rm eff}$ = 6000 K, as
well as log g $\le$ 2.00.  The sample analyzed here is dominated by stars with 
log g $\ge$ 2.00, T$_{\rm eff}$ = 4500 - 6000 K,
and near-solar metallicities, where the predicted effects on \ion{Fe}{1} are
less than 0.1 dex. 
More recent results presented in \citet{m10b} increase the number
of stars to five and indicate that LTE can be considered \textquotedblleft as good as non-LTE\textquotedblright\
for S$_{\rm H}$ $>$ 0.1 and metallicities between solar and -0.5 dex, based on
the analysis of stars such as Procyon and $\tau$ Cet.
With values of S$_{\rm H}\sim$ 0.1, combined with the points described above, 
LTE abundances from \ion{Fe}{1} and \ion{Fe}{2} are expected to be be very close to those derived from 
non-LTE (within hundreths of a dex) for the samples of stars with planets studied here.

\subsubsubsection {Observations of Fe in dwarfs and giants in clusters}

\begin{figure}
\epsscale{0.80}
\plotone{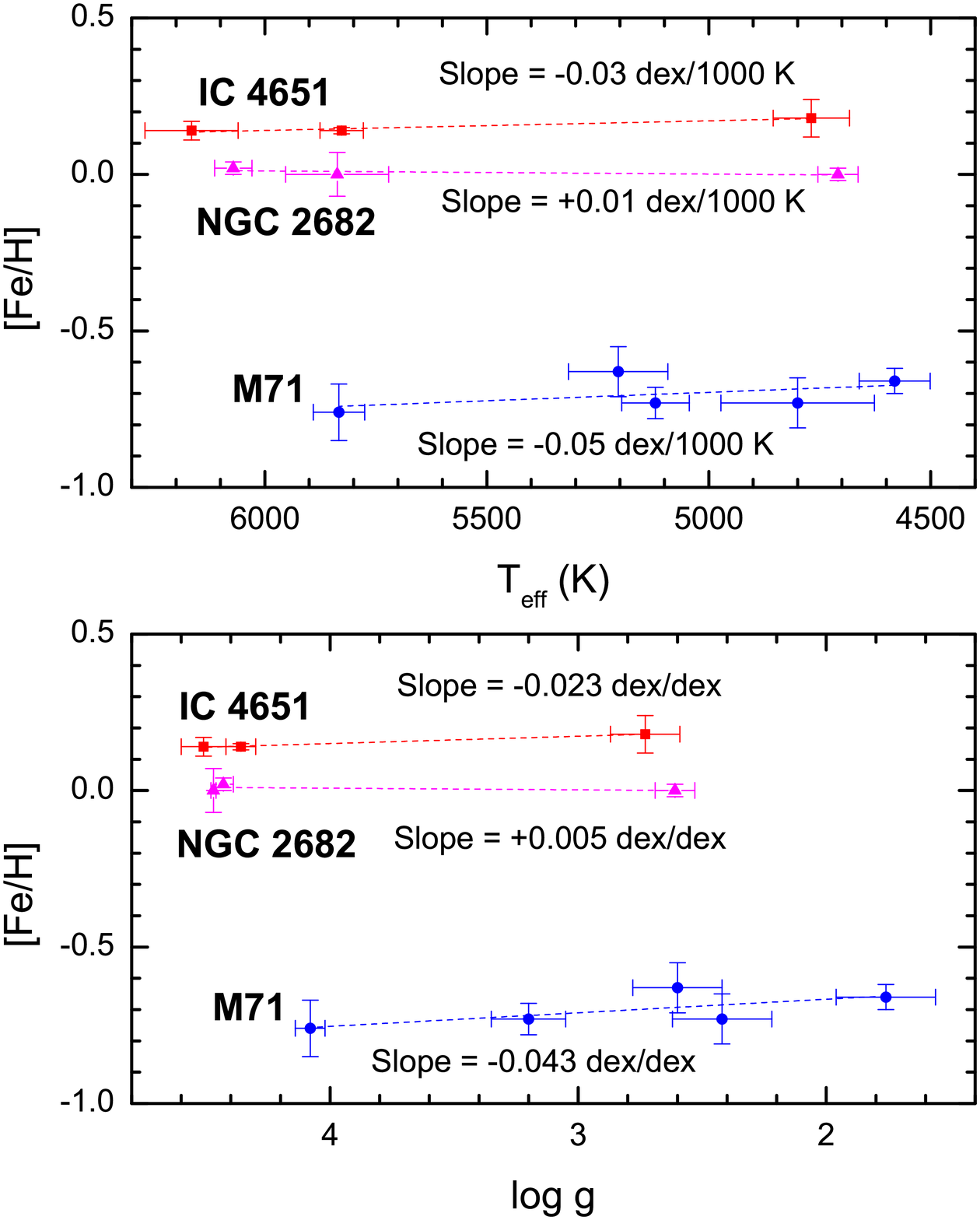}
\caption{Iron abundances versus effective temperatures (top panel) and surface gravities (bottom panel) 
for the globular cluster M71 (\citealt{r01}; blue circles) and the open clusters
NGC 2682 and IC 4651 (\citealt{s09}; respectively, magenta triangles and red squares). The dashed lines show 
the linear fit to the points for each case. No significant slopes are observed for any of the clusters.}
\label{clusters}
\end{figure}

The theoretically predicted small non-LTE effects on \ion{Fe}{1} described above
for near-solar metallicity stars are
born out by observations of real stars in clusters which contain uniform
Fe abundances.  One recent result relevant to this discussion is the abundance
analysis of giants, subgiants, and main-sequence stars from a number of
open clusters by \citet{s09}, as well as the results from the analysis of
main-sequence turn-off stars, subgiants, and giants in the globular cluster
M71 (one of the more metal-rich globular clusters) by \citet{r01}.
A short summary of these studies would note that no significant abundance
differences (i.e., $\Delta$[Fe/H] $\ge$ 0.05 dex) were found in LTE 
analyses of \ion{Fe}{1} lines between stars with
T$_{\rm eff}\sim$ 6000--6100 K and log g $\sim$ 4.2--4.6 when compared to
those with T$_{\rm eff}\sim$ 4500--4600 K and log g $\sim$ 1.7--2.5 in
any of the studied clusters.  Iron abundances from these studies are illustrated in Figure \ref{clusters}
for the globular cluster M71 (\citealt{r01}) and the open clusters
NGC 2682 and IC 4651 (\citealt{s09}); these particular clusters are
shown as the numbers of stars studied in each cluster were the largest,
and the clusters span a range in [Fe/H] overlapping that of the sample stars
included here.  The top panel shows Fe abundances 
(as [Fe/H]) plotted versus T$_{\rm eff}$ and the bottom panel is [Fe/H]
versus log g.  The points are average values for stars found along the
major phases of stellar evolution (main sequence, turn-off, subgiant and
giant branches), with the standard deviations in [Fe/H], T$_{\rm eff}$,
and log g shown for each sub-sample.  Linear least-squares fits were carried
out for each cluster and the derived slopes are labelled.  For the near-solar
metallicity open clusters, in particular, the slopes are very small and
not significant indicating a good agreement between the LTE Fe abundances
in dwarfs and giants.


Another piece of evidence that the Fe abundances of the 
stars analyzed here are not affected by significant non-LTE
effects is the comparison of stellar surface gravities
derived from the enforcement of LTE ionization equilibrium
between \ion{Fe}{1} and \ion{Fe}{2}, with surface gravities derived from
fitting stellar models to luminosities obtained from Hipparcos
parallaxes (so-called Hipparcos gravities), as illustrated
here in Figure \ref{loggs}. If Fe I and/or Fe II suffer from significant
departures from LTE, gravities may be adversely affected.
The mean difference in log g between these
methods for subgiants and giants is log g$_{\rm Hipp} -$ log g$_{\rm Spec}= -0.04\pm$0.12 dex. 
The standard deviation in this difference 
compares well with the expected uncertainty in defining log g
by any method, while the small offset of 0.04 dex indicates
excellent agreement between the two surface gravity methods.
This small difference indicates that there are not significant
departures from LTE populations in \ion{Fe}{1} and \ion{Fe}{2} in the
line-forming regions of near-solar metallicity dwarfs, subgiants,
and giants in the T$_{\rm eff}$ and log g regimes analyzed here.

As the analysis of all stars in the present study was done in a strictly
homogeneous manner, when coupled to the predictions that non-LTE corrections
to \ion{Fe}{1} in near-solar metallicity dwarfs and giants with the range of
stellar parameters studied here will be small (much less than 0.1 dex), and
the observations of rather uniform LTE iron abundances derived in
cluster dwarfs and giants, it is unlikely that non-LTE departures can explain the
differences observed here in the iron abundances of dwarfs and subgiants with planets, when
compared to giants with planets ($\sim$ 0.2 dex).

\subsubsection{Dwarfs, Subgiants, Giants, Dilution and Other Possibilities} 

The fact that the metallicity distribution of giant stars with planets in our sample is generally more metal poor
than the metallicity distribution found for the dwarfs cannot be explained 
in terms of Galactic chemical evolution, as these results are opposite from what would be expected
from chemical evolution: the giant sample stars being younger (on average) than the sample dwarfs would be 
more metal rich if one considers the effects of an age-metallicity relation.   
\citet{h09} propose, on the other hand, that the difference in the metallicity distributions of dwarfs and giants 
with planets is not related to the formation process of giants planets themselves, but results from a 
galactic effect instead. His conclusions are based on ages and metallicities of sample giants and dwarfs analyzed
by \citet{t07} and \citet{t08}. 
As radial mixing is a secular process, the sample of giants would be less contaminated by old, 
metal-rich wanderers of the inner disk. This scenario 
would only hold, however, if stars from the inner disk have a higher percentage of giant planets 
than stars born at the solar radius and assumes a metallicity gradient for the Galactic disk. 

A relevant question concerning metallicity distributions of planet hosting stars in
different evolutionary stages connects the possibility of late accretion of metal rich material onto the star
to the dilution of this abundance signature as the star develops a deeper convective envelope. 
The expectation in such a scenario would be that the metal rich signature which is due to accretion
would vanish as stars become giants; their convective zones become larger and the metal rich material 
becomes diluted. If the high metallicity observed for the main-sequence stars hosting giant planets 
is indeed restricted to the outer envelope it is expected that subgiants will have a systematically
lower metallicity than the dwarfs.

\begin{figure}
\epsscale{1.00}
\plotone{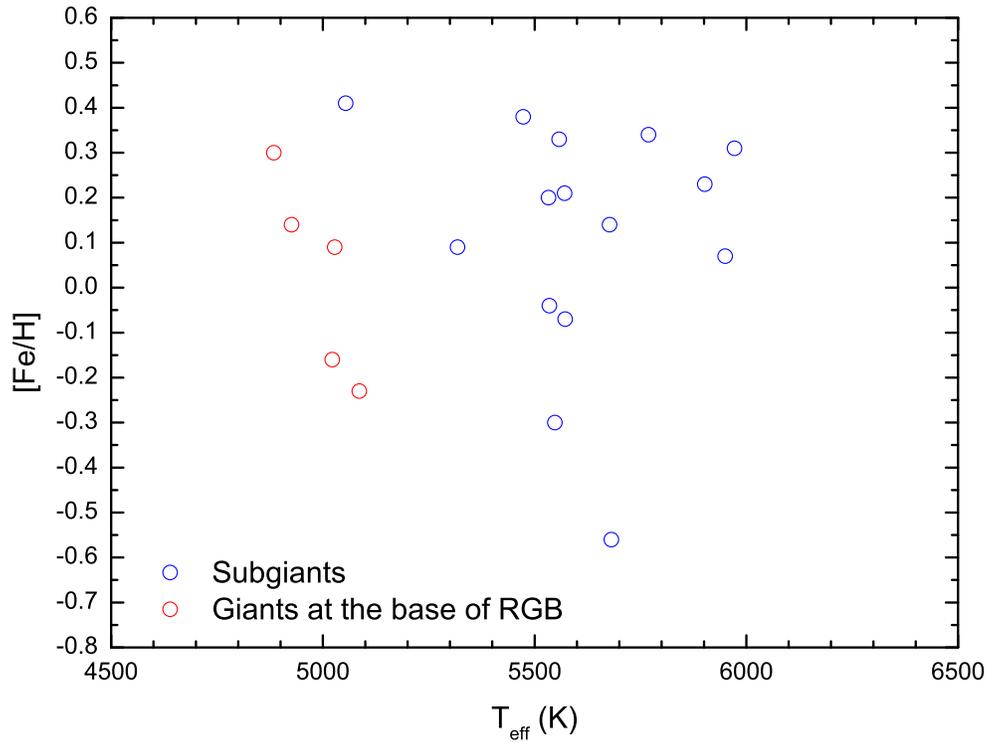}
\caption{Metallicities versus effective temperatures for the studied sample of subgiant stars hosting planets 
(blue open circles). The absence of a trend in this figure indicates the stars are not experiencing 
increasing dilution of their convective zones. Sample giants (red open circles) which are at the base of 
the red giant branch ($M_{bol} \simeq 2.82$) are also shown for comparison.}
\label{feh_versus_teff}
\end{figure}

Taken at face value, the metallicity distributions of planet hosting dwarfs, subgiants and giants 
obtained in this study are not in line with the dilution picture as there is not a consistent decrease in the average 
metallicities for planet-hosting stars going from dwarfs (+0.11 dex) to subgiants (+0.12 dex), to giants (-0.06 dex);
in particular between the dwarfs and subgiants. 
In addition, the absence of a trend in the plot of effective temperature versus stellar
metallicity for sample subgiants (shown in Figure \ref{feh_versus_teff}) indicates the absence of dilution on the subgiant branch. 
A trend in the run of metallicity with effective temperature would be expected if the stars experienced 
increased dilution as they evolve redward on the subgiant branch, but this gradient is flat.  
The giants in our sample which are at the base of the red giant branch (with $M_{bol} \simeq 2.82$; see Figure \ref{Mbol_teff}), are also shown 
in Figure \ref{feh_versus_teff} as it is at this stage that the convective zone deepens significantly. 
Note that a similar range in metallicity (roughly between [Fe/H]= -0.3 to +0.3 dex) 
is encompassed by the subgiants and giants in the figure, which indicates no significant differences between the
metallicities of subgiants and giants at the base of the RGB. 
This result is in agreement with the findings of \citet{fv05} who also
do not find a metallicity gradient as a function of $T_{eff}$ along the subgiant branch
and conclude that subgiants do not exhibit any evidence for dilution \citep[see also][] {j10a}. 


Without evidence for dilution along the subgiant branch, the observations point to
a scenario to explain the more metal poor distribution observed for giant stars 
in comparison with dwarfs which is related to the fact that the higher masses of the giant stars 
compensate for the lower metallicities by allowing, or favoring, the formation of planets 
because higher mass stars have on average  disks with larger masses. 
A number of studies (e.g. \citealt{n00}) find that disk mass increases
with stellar mass. As disk mass increases, the surface density ($\sigma$) within typical
protoplanetary disks also increases and larger values of $\sigma$ favor the formation
of giant planets in the core accretion model of planetary formation (\citealt{il04}; \citealt{l04}; see also \citealt{j10a}). 
The precipitation of a substantial planetary core which begins to accrete gas requires a threshold density
of solid material (which consists of the heavier elements, or metals).

The observation that the average metallicity of planet-hosting stars is related to the average
mass within a stellar sample, with giants representing the more massive but lower metallicity
population, is taken as an observational signature of core-accretion as the main mechanism
for planetary formation, at least for planets which form relatively close to their parent
stars. 
The disk instability mechanism (see review by \citealt{boss10}) may still be important for the massive
planets which form at large distances from their parent stars.  Examples of such systems
may be the recently imaged planets found around HR 8799 (\citealt{m08}) and Fomalhaut
(\citealt{k08}). The nature of planetary system architectures is quite likely a function of both
stellar mass and metallicity.


\section{Conclusions}

\label{conc}

It is now well established that stars hosting giant planets have on average higher
metallicities than stars which do not host closely orbiting giant planets 
(see, e.g. \citealt{g06}; \citealt{us07}; \citealt{v10} for reviews). So far, however, most of the
studies have concentrated on host stars which are on the main sequence. 
Such a finding was recently corroborated from metallicities obtained in the homogeneous analysis of a 
large sample of main-sequence planet hosts and a control sample of stars without closely
orbiting giant planets, which was presented in \citetalias{ghezzi10}. 
The present study extends the main-sequence sample in \citetalias{ghezzi10} by adding planet hosting 
stars which are evolved from the main sequence.

We have determined stellar parameters and metallicities for a sample of 15 subgiants
and 16 giants with planets discovered via radial velocity surveys and 14
comparison subgiants which have been found to exhibit nearly
constant radial velocities and are not likely to host
large, closely orbiting planets. The stellar parameters and iron abundances
were derived from a classical spectroscopic analysis (similar to \citetalias{ghezzi10}). 

Our results are summarized as follows:

1) One strong point of the present study is the strictly homogeneous abundance 
analyses performed for the samples of dwarfs (\citetalias{ghezzi10}), subgiants and giants.
An additional important aspect is the sample of disk subgiant stars
which are known to be RV stable (\citealt{fv05}) and can be used as 
a comparison sample for the subgiant planet-hosting stars.

2) The subgiant sample in this study is found to be a slightly older population which has evolved mostly
from the same underlying population as the dwarfs analyzed in \citetalias{ghezzi10};
the sample dwarfs and subgiants have significant overlap in their mass ranges. 
The sample giants, however, are evolved from stars which 
are more massive and are, on average, the youngest of all studied targets.

3) The metallicity distribution obtained for our sample of 16 giant planet-hosting stars 
displays an average that is more metal poor by 0.17 dex than the metallicity distribution 
obtained in \citetalias{ghezzi10} for the sample of planet-hosting dwarfs (N=117). 
When literature iron abundance results for all other presently known giant planet-hosting stars 
are included, in order to improve the giant star statistics to a total of 37 stars, the offset in the average 
metallicities between dwarf and giant planet-hosting stars is confirmed and becomes
marginally larger (0.23 dex). 

4) The average metallicity of the planet-hosting subgiant sample is metal-rich
relative to the Sun, $\langle [Fe/H]\rangle$ = +0.12 dex. The latter distribution is similar to that
obtained for the planet hosting dwarf sample, and on average more metal rich than that 
of subgiants without planets by 0.21 dex. This abundance difference between the subgiants with and 
without planets is in general agreement, within the uncertainties, 
with the abundance shift that is found for dwarfs with and without planets. 

5) The absence of a trend in the derived iron abundances with effective temperature for the
sample subgiant stars shows no evidence for dilution on the subgiant branch.
This flat gradient plus the fact that there is not a significant difference between
the metallicity distributions of subgiant planet-hosting stars in comparison to 
dwarf planet-hosting stars, as would be expected from the more extended convective
envelopes of subgiants in comparison with dwarfs, weakens the possibility of dilution as 
a viable explanation for the lower metallicity found for the giant stars. 

6) In the absence of substantial evidence for the dilution of accreted metal-rich material, the results in
this study favor a scenario to explain the more metal poor distribution observed for giant stars
in comparison with that of dwarfs which is related to the fact that the higher masses of the
giant stars compensate for the lower metallicities, as higher mass stars have on average
more massive disks with more metals available for planet formation through core accretion.


\acknowledgements{We thank Leo Girardi for helping with the computation of the evolutionary
parameters. K.C. thanks Andreas Korn and Ivan Hubeny for discussions concerning non-LTE effects
and Joan Najita and Luca Pasquini for fruitful discussions. L.G. thanks Oliver Sch\"{u}tz for his 
valuable help in using the FEROS DRS package and Cl\'audio Bastos for conducting the observations for 
target star HD 114613.L.G. acknowledges financial support by CNPq. 
Research presented here was supported in-part by NASA grant NNH08AJ581.}


\begin{deluxetable}{lccccc}
\tablecolumns{6}
\tablewidth{0pc}
\tablecaption{Log of Observations\label{obslog}}
\tablehead{
\colhead{Star} & \colhead{$V$} & \colhead{Observation} & \colhead{$T_{exp}$} & \colhead{S/N} & \colhead{Classification} \\
\colhead{} & \colhead{} & \colhead{Date} & \colhead{(s)} & \colhead{($\sim$ 6700 \AA)} & \colhead{}}
\startdata
\sidehead{\textit{Planet Hosting Stars}}
HD 5319         &  8.05    & 2007 Aug 28    &  1200    & 282 &  G \\
HD 10697        &  6.27    & 2007 Aug 29    &   200    & 277 & SG \\
HD 11977        &  4.68    & 2007 Aug 30    &    80    & 344 &  G \\
HD 11964        &  6.42    & 2007 Aug 30    &   200    & 313 & SG \\
HD 16400        &  5.65    & 2008 Aug 19    &   200    & 352 &  G \\
HD 23127        &  8.58    & 2007 Aug 30    &  1800    & 291 & SG \\
HD 27442        &  4.44    & 2007 Oct 02    &    15    & 138 &  G \\
HD 28305        &  3.53    & 2007 Aug 30    &    30    & 291 &  G \\
HD 33283        &  8.05    & 2007 Aug 30    &  1200    & 385 & SG \\
HD 38529        &  5.95    & 2007 Oct 02    &   200    & 314 & SG \\
HD 47536        &  5.25    & 2007 Apr 08    &    80    & 369 &  G \\
HD 59686        &  5.45    & 2007 Apr 08    &    80    & 217 &  G \\
NGC 2423 3      & 10.04    & 2007 Aug 28    &  3000    & 183 &  G \\
HD 73526        &  8.99    & 2007 Apr 08    &  3000    & 301 & SG \\
HD 88133        &  8.01    & 2007 Apr 07    &  1200    & 290 & SG \\
NGC 4349 127    & 10.83    & 2008 Apr 06    & 10800    & 196 &  G \\
HD 117176       &  4.97    & 2007 Apr 06    &    80    & 415 & SG \\
HD 122430       &  5.47    & 2007 Apr 06    &    80    & 199 &  G \\
HD 154857       &  7.24    & 2007 Apr 06    &   480    & 439 & SG \\
HD 156846       &  6.50    & 2008 Apr 06    &   500    & 417 & SG \\
HD 159868       &  7.24    & 2007 Apr 06    &   480    & 460 & SG \\
HD 171028       &  8.31    & 2007 Aug 28    &  1200    & 331 & SG \\
HD 175541       &  8.02    & 2007 Aug 28    &  1200    & 291 &  G \\
HD 177830       &  7.18    & 2007 Aug 29    &   480    & 207 & SG \\
HD 188310       &  4.71    & 2008 Apr 06    &   100    & 344 &  G \\
HD 190647       &  7.78    & 2007 Aug 28    &  1200    & 363 & SG \\
HD 192699       &  6.44    & 2007 Aug 28    &   200    & 259 &  G \\
HD 199665       &  5.51    & 2008 Apr 06    &   200    & 320 &  G \\
HD 210702       &  5.93    & 2007 Aug 28    &   200    & 280 &  G \\
HD 219449       &  4.24    & 2007 Aug 30    &    60    & 316 &  G \\
HD 224693       &  8.23    & 2007 Aug 29    &  1200    & 336 & SG \\
\sidehead{\textit{Control Sample}}
HD 2151         &  2.82    & 2008 Aug 20    &    15    & 363 & SG \\
HD 18907        &  5.88    & 2008 Aug 20    &   200    & 407 & SG \\
HD 33473        &  6.75    & 2008 Aug 20    &   500    & 441 & SG \\
HD 114613       &  4.85    & 2008 Feb 21    &   100    & 446 & SG \\
HD 121384       &  6.00    & 2008 Apr 07    &   200    & 414 & SG \\
HD 140785       &  7.38    & 2008 Apr 07    &   500    & 371 & SG \\
HD 168060       &  7.34    & 2008 Aug 20    &   500    & 331 & SG \\
HD 168723       &  3.23    & 2008 Aug 20    &    15    & 326 &  G \\
HD 188641       &  7.34    & 2008 Aug 19    &   500    & 409 & SG \\
HD 196378       &  5.11    & 2008 Aug 19    &   100    & 433 & SG \\
HD 205420       &  6.45    & 2008 Aug 19    &   200    & 307 & SG \\
HD 208801       &  6.24    & 2008 Aug 19    &   200    & 301 & SG \\
HD 212330       &  5.31    & 2008 Aug 20    &   100    & 273 & SG \\
HD 219077       &  6.12    & 2008 Aug 19    &   200    & 372 & SG \\
HD 221420       &  5.82    & 2008 Aug 20    &   200    & 307 & SG \\
\enddata
\tablecomments{SG = Subgiant; G = Giant.}
\end{deluxetable}


\begin{deluxetable}{lccccccccr}
\tablecolumns{10}
\tablewidth{0pc}
\tablecaption{Atmospheric Parameters and Metallicities\label{atm_par}}
\tablehead{
\colhead{Star} & \colhead{T$_{eff}$} & \colhead{$\log$ g} & \colhead{$\xi$} & \colhead{A(Fe)} & \colhead{$\sigma$} & \colhead{N} & \colhead{$\sigma$} & \colhead{N} & \colhead{[Fe/H]} \\
\colhead{} & \colhead{(K)} & \colhead{} & \colhead{(km $\rm s^{-1}$)} & \colhead{} & \colhead{(\ion{Fe}{1})} & \colhead{(\ion{Fe}{1})} & \colhead{(\ion{Fe}{2})} & \colhead{(\ion{Fe}{2})} & \colhead{}}
\startdata
\sidehead{\textit{Planet Hosting Stars}}
HD 5319         & 4926    & 3.33     & 1.17     & 7.57     & 0.10    & 23    & 0.06    &  9     &  0.14 \\
HD 10697        & 5677    & 4.06     & 1.28     & 7.57     & 0.08    & 27    & 0.05    & 12     &  0.14 \\
HD 11977        & 4972    & 2.64     & 1.42     & 7.27     & 0.08    & 25    & 0.07    & 11     & -0.16 \\
HD 11964        & 5318    & 3.77     & 1.12     & 7.52     & 0.07    & 27    & 0.05    & 11     &  0.09 \\
HD 16400        & 4783    & 2.39     & 1.46     & 7.36     & 0.11    & 25    & 0.09    &  7     & -0.07 \\
HD 23127        & 5769    & 4.01     & 1.30     & 7.77     & 0.09    & 27    & 0.06    & 10     &  0.34 \\
HD 27442        & 4884    & 3.39     & 1.31     & 7.73     & 0.12    & 22    & 0.07    &  6     &  0.30 \\
HD 28305        & 4963    & 2.87     & 1.68     & 7.60     & 0.13    & 25    & 0.07    &  7     &  0.17 \\
HD 33283        & 5972    & 4.02     & 1.42     & 7.74     & 0.08    & 27    & 0.06    & 12     &  0.31 \\
HD 38529        & 5558    & 3.62     & 1.32     & 7.76     & 0.09    & 26    & 0.06    & 10     &  0.33 \\
HD 47536        & 4588    & 2.17     & 2.03     & 6.82     & 0.10    & 23    & 0.04    &  4     & -0.61 \\
HD 59686        & 4740    & 2.66     & 1.58     & 7.57     & 0.14    & 22    & 0.04    &  5     &  0.14 \\
NGC 2423 3      & 4680    & 2.55     & 1.67     & 7.43     & 0.13    & 25    & 0.05    &  8     &  0.00 \\
HD 73526        & 5571    & 3.89     & 1.15     & 7.64     & 0.07    & 26    & 0.05    & 10     &  0.21 \\
HD 88133        & 5473    & 3.94     & 1.09     & 7.81     & 0.07    & 24    & 0.04    &  8     &  0.38 \\
NGC 4349 127    & 4519    & 1.92     & 2.08     & 7.22     & 0.12    & 17    & 0.08    &  6     & -0.21 \\
HD 117176       & 5535    & 3.98     & 1.12     & 7.39     & 0.07    & 27    & 0.06    & 12     & -0.04 \\
HD 122430       & 4367    & 1.71     & 1.71     & 7.27     & 0.14    & 17    & 0.08    &  7     & -0.16 \\
HD 154857       & 5548    & 3.82     & 1.34     & 7.13     & 0.07    & 26    & 0.05    & 12     & -0.30 \\
HD 156846       & 5950    & 3.84     & 1.62     & 7.50     & 0.09    & 24    & 0.07    & 11     &  0.07 \\
HD 159868       & 5572    & 3.90     & 1.21     & 7.36     & 0.06    & 26    & 0.06    & 12     & -0.07 \\
HD 171028       & 5681    & 3.88     & 1.71     & 6.87     & 0.06    & 24    & 0.06    & 10     & -0.56 \\
HD 175541       & 5022    & 3.19     & 1.15     & 7.27     & 0.05    & 24    & 0.06    & 11     & -0.16 \\
HD 177830       & 5054    & 3.83     & 1.30     & 7.84     & 0.11    & 20    & 0.05    &  6     &  0.41 \\
HD 188310       & 4783    & 2.66     & 1.57     & 7.30     & 0.12    & 23    & 0.07    &  7     & -0.13 \\
HD 190647       & 5533    & 3.92     & 1.12     & 7.63     & 0.06    & 25    & 0.04    & 10     &  0.20 \\
HD 192699       & 5086    & 3.18     & 1.17     & 7.20     & 0.07    & 25    & 0.06    & 11     & -0.23 \\
HD 199665       & 4948    & 2.69     & 1.31     & 7.34     & 0.09    & 24    & 0.09    &  8     & -0.09 \\
HD 210702       & 5028    & 3.40     & 1.24     & 7.52     & 0.09    & 26    & 0.05    &  7     &  0.09 \\
HD 219449       & 4812    & 2.78     & 1.72     & 7.48     & 0.11    & 23    & 0.06    &  5     &  0.05 \\
HD 224693       & 5902    & 3.97     & 1.36     & 7.66     & 0.10    & 27    & 0.06    & 12     &  0.23 \\
\sidehead{\textit{Control Sample}}
HD 2151         & 5866    & 4.00     & 1.51     & 7.32     & 0.08    & 24    & 0.06    & 10     & -0.11 \\
HD 18907        & 5212    & 3.92     & 1.20     & 6.87     & 0.06    & 24    & 0.04    & 10     & -0.56 \\
HD 33473        & 5608    & 3.60     & 1.36     & 7.21     & 0.08    & 26    & 0.05    & 12     & -0.22 \\
HD 114613       & 5717    & 3.92     & 1.30     & 7.61     & 0.07    & 26    & 0.07    & 12     &  0.18 \\
HD 121384       & 5249    & 3.67     & 1.24     & 6.93     & 0.07    & 27    & 0.05    & 12     & -0.50 \\
HD 140785       & 5723    & 3.98     & 1.18     & 7.40     & 0.05    & 23    & 0.04    & 12     & -0.03 \\
HD 168060       & 5577    & 3.93     & 1.14     & 7.72     & 0.09    & 27    & 0.04    & 10     &  0.29 \\
HD 168723       & 4944    & 3.12     & 1.25     & 7.26     & 0.08    & 26    & 0.04    &  8     & -0.17 \\
HD 188641       & 5816    & 3.98     & 1.37     & 7.31     & 0.06    & 24    & 0.05    & 12     & -0.12 \\
HD 196378       & 5996    & 3.92     & 1.78     & 6.99     & 0.04    & 20    & 0.05    & 10     & -0.44 \\
HD 205420       & 6255    & 3.89     & 1.99     & 7.43     & 0.06    & 20    & 0.07    & 12     &  0.00 \\
HD 208801       & 5061    & 3.80     & 1.08     & 7.59     & 0.08    & 21    & 0.06    &  8     &  0.16 \\
HD 212330       & 5670    & 3.91     & 1.33     & 7.41     & 0.07    & 27    & 0.05    & 12     & -0.02 \\
HD 219077       & 5321    & 3.80     & 1.13     & 7.27     & 0.08    & 27    & 0.04    & 11     & -0.16 \\
HD 221420       & 5899    & 4.04     & 1.48     & 7.77     & 0.05    & 20    & 0.06    & 12     &  0.34 \\
\enddata
\end{deluxetable}


\begin{deluxetable}{lccccccccccccc}
\tabletypesize{\footnotesize}
\tablecolumns{14}
\rotate
\tablewidth{0pc}
\tablecaption{Evolutionary Parameters.\label{evol_par}}
\tablehead{
\colhead{Star} & \colhead{$\pi$} & \colhead{$\sigma_{\pi}$} & \colhead{$A_{V}$} & \colhead{$\log(L/L_{\sun})$} & \colhead{$\sigma_{\log(L/L_{\sun})}$} & \colhead{$R$} & \colhead{$\sigma_{R}$} & \colhead{$M$} & \colhead{$\sigma(M)$} & \colhead{$\log g_{Hipp}$} & \colhead{$\sigma(\log g_{Hipp})$} & \colhead{Age} & \colhead{$\sigma$(Age)} \\
\colhead{} & \colhead{(mas)} & \colhead{(mas)} & \colhead{(mag)} & \colhead{} & \colhead{} & \colhead{(R$_{\sun}$)} & \colhead{(R$_{\sun}$)} & \colhead{(M$_{\sun}$)} & \colhead{(M$_{\sun}$)} & \colhead{} & \colhead{} & \colhead{(Gyr)} & \colhead{(Gyr)}}
\startdata
\sidehead{\textit{Planet Hosting Stars}}
HD 5319          &   8.74     & 0.86     & 0.10     & 0.952     & 0.105     &  3.97     & 0.43     & 1.40     & 0.14     & 3.35     & 0.10     & 3.30     & 1.11 \\
HD 10697         &  30.70     & 0.43     & 0.00     & 0.446     & 0.061     &  1.69     & 0.06     & 1.11     & 0.03     & 3.99     & 0.03     & 6.75     & 0.71 \\
HD 11977         &  14.91     & 0.16     & 0.16     & 1.849     & 0.061     & 11.04     & 0.43     & 2.27     & 0.29     & 2.68     & 0.07     & 0.83     & 0.27 \\
HD 11964         &  30.44     & 0.60     & 0.10     & 0.466     & 0.062     &  1.97     & 0.08     & 1.12     & 0.03     & 3.86     & 0.03     & 7.02     & 0.67 \\
HD 16400         &  10.81     & 0.45     & 0.08     & 1.741     & 0.074     & 10.50     & 0.45     & 1.43     & 0.31     & 2.52     & 0.11     & 2.66     & 1.46 \\
HD 23127         &  10.13     & 0.67     & 0.21     & 0.559     & 0.083     &  1.81     & 0.13     & 1.21     & 0.05     & 3.97     & 0.06     & 4.66     & 0.81 \\
HD 27442         &  54.83     & 0.15     & 0.03     & 0.782     & 0.060     &  3.50     & 0.15     & 1.35     & 0.08     & 3.44     & 0.06     & 3.79     & 0.85 \\
HD 28305         &  22.24     & 0.25     & 0.06     & 1.924     & 0.061     & 12.69     & 0.46     & 2.75     & 0.11     & 2.64     & 0.03     & 0.51     & 0.09 \\
HD 33283         &  10.62     & 0.62     & 0.21     & 0.718     & 0.079     &  2.08     & 0.13     & 1.39     & 0.06     & 3.91     & 0.05     & 2.93     & 0.41 \\
HD 38529         &  25.46     & 0.40     & 0.03     & 0.754     & 0.062     &  2.49     & 0.10     & 1.37     & 0.02     & 3.74     & 0.03     & 3.35     & 0.14 \\
HD 47536         &   8.11     & 0.23     & 0.11     & 2.204     & 0.065     & 19.84     & 1.09     & 1.15     & 0.25     & 1.87     & 0.12     & 4.38     & 2.58 \\
HD 59686         &  10.32     & 0.28     & 0.00     & 1.840     & 0.064     & 11.80     & 0.60     & 2.27     & 0.30     & 2.62     & 0.09     & 0.92     & 0.33 \\
NGC 2423 3       &   1.31     & 0.03     & 0.39     & 1.966     & 0.099     & 14.11     & 0.88     & 2.16     & 0.38     & 2.44     & 0.11     & 0.96     & 0.40 \\
HD 73526         &   9.93     & 1.01     & 0.06     & 0.369     & 0.107     &  1.53     & 0.16     & 1.05     & 0.05     & 4.05     & 0.08     & 8.50     & 1.34 \\
HD 88133         &  12.28     & 0.88     & 0.04     & 0.576     & 0.087     &  2.04     & 0.15     & 1.20     & 0.06     & 3.87     & 0.04     & 5.22     & 0.90 \\
NGC 4349 127     &   0.45     & 0.01     & 1.08     & 2.889     & 0.205     & 44.72     & 2.46     & 3.77     & 0.36     & 1.68     & 0.07     & 0.20     & 0.05 \\
HD 117176        &  55.60     & 0.24     & 0.01     & 0.467     & 0.060     &  1.83     & 0.06     & 1.08     & 0.03     & 3.91     & 0.03     & 7.83     & 0.63 \\
HD 122430        &   7.42     & 0.33     & 0.28     & 2.325     & 0.071     & 24.49     & 1.78     & 1.53     & 0.31     & 1.81     & 0.12     & 2.18     & 1.11 \\
HD 154857        &  15.57     & 0.71     & 0.12     & 0.712     & 0.075     &  2.40     & 0.14     & 1.21     & 0.06     & 3.73     & 0.04     & 4.43     & 0.63 \\
HD 156846        &  21.00     & 0.51     & 0.14     & 0.724     & 0.067     &  2.11     & 0.08     & 1.36     & 0.06     & 3.89     & 0.04     & 3.17     & 0.47 \\
HD 159868        &  17.04     & 0.76     & 0.08     & 0.612     & 0.072     &  2.11     & 0.11     & 1.19     & 0.04     & 3.83     & 0.04     & 5.31     & 0.76 \\
HD 171028        &  11.10     & 1.85     & 0.31     & 0.651     & 0.165     &  2.06     & 0.28     & 1.03     & 0.09     & 3.79     & 0.08     & 7.25     & 2.44 \\
HD 175541        &   7.87     & 0.95     & 0.32     & 1.126     & 0.127     &  4.55     & 0.57     & 1.37     & 0.16     & 3.23     & 0.11     & 3.11     & 1.16 \\
HD 177830        &  16.94     & 0.63     & 0.08     & 0.698     & 0.068     &  2.85     & 0.16     & 1.41     & 0.03     & 3.65     & 0.04     & 3.14     & 0.20 \\
HD 188310        &  17.77     & 0.29     & 0.09     & 1.690     & 0.062     & 10.23     & 0.39     & 1.16     & 0.28     & 2.45     & 0.11     & 4.63     & 2.88 \\
HD 190647        &  17.46     & 0.81     & 0.15     & 0.402     & 0.076     &  1.68     & 0.09     & 1.07     & 0.03     & 3.99     & 0.04     & 7.96     & 0.81 \\
HD 192699        &  15.24     & 0.57     & 0.04     & 1.064     & 0.068     &  4.41     & 0.23     & 1.38     & 0.13     & 3.26     & 0.07     & 2.90     & 0.88 \\
HD 199665        &  13.28     & 0.31     & 0.05     & 1.578     & 0.063     &  8.29     & 0.31     & 2.01     & 0.10     & 2.87     & 0.04     & 1.10     & 0.16 \\
HD 210702        &  18.20     & 0.39     & 0.05     & 1.125     & 0.063     &  4.83     & 0.24     & 1.72     & 0.13     & 3.27     & 0.06     & 1.68     & 0.36 \\
HD 219449        &  21.77     & 0.29     & 0.10     & 1.702     & 0.061     & 10.16     & 0.45     & 1.74     & 0.35     & 2.63     & 0.11     & 1.69     & 0.81 \\
HD 224693        &  10.16     & 0.91     & 0.10     & 0.646     & 0.101     &  1.90     & 0.17     & 1.30     & 0.08     & 3.96     & 0.06     & 3.54     & 0.68 \\
\sidehead{\textit{Control Sample}}
HD 2151          & 134.07     & 0.11     & 0.02     & 0.547     & 0.060     &  1.77     & 0.05     & 1.13     & 0.04     & 3.96     & 0.03     & 6.13     & 0.88 \\
HD 18907         &  31.06     & 0.36     & 0.09     & 0.678     & 0.061     &  2.66     & 0.10     & 1.02     & 0.06     & 3.56     & 0.05     & 7.62     & 1.51 \\
HD 33473         &  18.69     & 0.49     & 0.09     & 0.731     & 0.064     &  2.41     & 0.10     & 1.25     & 0.04     & 3.74     & 0.03     & 4.06     & 0.37 \\
HD 114613        &  48.38     & 0.29     & 0.04     & 0.631     & 0.060     &  2.06     & 0.07     & 1.26     & 0.03     & 3.87     & 0.03     & 4.22     & 0.27 \\
HD 121384        &  25.84     & 0.48     & 0.07     & 0.776     & 0.062     &  2.95     & 0.12     & 1.15     & 0.08     & 3.53     & 0.06     & 4.90     & 1.21 \\
HD 140785        &  17.54     & 0.56     & 0.12     & 0.536     & 0.070     &  1.85     & 0.08     & 1.13     & 0.03     & 3.92     & 0.04     & 6.33     & 0.75 \\
HD 168060        &  21.07     & 0.65     & 0.06     & 0.375     & 0.066     &  1.61     & 0.07     & 1.07     & 0.02     & 4.02     & 0.04     & 7.92     & 0.56 \\
HD 168723        &  53.93     & 0.18     & 0.06     & 1.279     & 0.060     &  6.00     & 0.24     & 1.41     & 0.16     & 3.00     & 0.07     & 2.81     & 1.01 \\
HD 188641        &  16.14     & 0.82     & 0.16     & 0.637     & 0.078     &  1.98     & 0.12     & 1.18     & 0.04     & 3.88     & 0.04     & 5.30     & 0.75 \\
HD 196378        &  40.55     & 0.27     & 0.05     & 0.684     & 0.060     &  2.01     & 0.06     & 1.10     & 0.03     & 3.84     & 0.03     & 5.62     & 0.43 \\
HD 205420        &  15.81     & 0.39     & 0.04     & 0.939     & 0.064     &  2.47     & 0.09     & 1.53     & 0.05     & 3.80     & 0.03     & 2.09     & 0.22 \\
HD 208801        &  27.11     & 0.41     & 0.05     & 0.653     & 0.061     &  2.73     & 0.13     & 1.34     & 0.05     & 3.66     & 0.04     & 3.61     & 0.41 \\
HD 212330        &  48.63     & 0.34     & 0.04     & 0.449     & 0.060     &  1.71     & 0.06     & 1.07     & 0.03     & 3.97     & 0.03     & 7.78     & 0.80 \\
HD 219077        &  34.07     & 0.37     & 0.05     & 0.470     & 0.061     &  1.99     & 0.08     & 1.06     & 0.04     & 3.83     & 0.03     & 8.27     & 0.70 \\
HD 221420        &  31.81     & 0.27     & 0.07     & 0.605     & 0.060     &  1.88     & 0.05     & 1.31     & 0.04     & 3.97     & 0.03     & 3.43     & 0.47 \\
\enddata
\end{deluxetable}



\begin{deluxetable}{lrl}
\tablecolumns{3}
\tablewidth{0pc}
\tablecaption{Giant Stars Comparison with the Literature\label{giant_lit}}
\tablehead{
\colhead{Star} & \colhead{[Fe/H]} & \colhead{Reference}}
\startdata
\sidehead{\textit{Results from This Work}}
HD 5319      &  0.14 & This study  \\
             &  0.15 & \citet{vf05}  \\
HD 11977     & -0.16 & This study \\
             & -0.21 & \citet{dS06} \\
             & -0.09 & \citet{s06} \\
HD 16400     & -0.07 & This study \\
             & -0.06 & \citet{t08} \\
HD 27442     &  0.30 & This study \\
             &  0.42 & \citet{s03} \\
             &  0.42 & \citet{vf05} \\
HD 28305     &  0.17 & This study \\
             &  0.11 & \citet{m06} \\
             &  0.20 & \citet{schuler06} \\
             &  0.17 & \citet{sato07} \\
             &  0.05 & \citet{hm07} \\
             &  0.13 & \citet{t08} \\
HD 47536     & -0.61 & This study \\
             & -0.54 & \citet{s04} \\
             & -0.54 & \citet{sadakane05} \\
             & -0.68 & \citet{dS06} \\
HD 59686     &  0.14 & This study \\
             &  0.28 & \citet{s05} \\
             &  0.11 & \citet{sadakane05} \\
             &  0.02 & \citet{m06} \\
             &  0.15 & \citet{hm07} \\
NGC 2423 3   &  0.00 & This study \\
             &  0.00 & \citet{s09} \\
NGC 4349 127 & -0.21 & This study \\
             & -0.14 & \citet{s09}\\
HD 122430    & -0.16 & This study \\
             & -0.05 & \citet{dS06} \\
HD 175541    & -0.16 & This study \\
             & -0.07 & \citet{vf05} \\
             & -0.07 & \citet{j07} \\
HD 188310    & -0.13 & This study \\
             & -0.21 & \citet{sato08a} \\
             & -0.18 & \citet{t08} \\
HD 192699    & -0.23 & This study \\
             & -0.15 & \citet{j07} \\
HD 199665    & -0.09 & This study \\
             & -0.05 & \citet{sato08a} \\
             & -0.05 & \citet{t08} \\
HD 210702    &  0.09 & This study \\
             &  0.06 & \citet{lh07} \\
             & +0.12 & \citet{j07} \\
HD 219449    &  0.05 & This study \\
             &  0.05 & \citet{s05} \\
             &  0.09 & \citet{sadakane05} \\
             &  0.05 & \citet{lh07} \\
             & -0.03 & \citet{hm07}\\
\sidehead{\textit{Literature Results}}
HD 13189     & -0.58 & \citet{schuler05} \\
             & -0.39 & \citet{s06} \\
             & -0.49 & Average \\
HD 17092     &  0.22 & \citet{n07}  \\
HD 32518     & -0.15 & \citet{d09a} \\
HD 62509     &  0.05 & \citet{sadakane05} \\
             &  0.19 & \citet{h06} \\
             &  0.17 & \citet{lh07} \\
             &  0.07 & \citet{hm07} \\
             &  0.12 & Average \\
4 UMa        & -0.16 & \citet{lh07} \\         
             & -0.25 & \citet{d07} \\
             & -0.21 & Average \\
HD 81688     & -0.36 & \citet{sato08a} \\
BD+20 2457   & -1.00 & \citet{n09b} \\
gamma 1 Leo  & -0.49 & \citet{mw90} \\              
             & -0.51 & \citet{h10} \\
             & -0.50 & Average \\
HD 102272    & -0.26 & \citet{n09a} \\
HD 104985    & -0.35 & \citet{sato03} \\
             & -0.28 & \citet{s05} \\
             & -0.15 & \citet{t05} \\
             & -0.26 & \citet{lh07} \\
             & -0.26 & Average \\
HD 110014    &  0.19 & \citet{dS06} \\
11 UMi       &  0.04 & \citet{d09a} \\   
HIP 75458    &  0.03 & \citet{mw90} \\              
             &  0.09 & \citet{s03} \\
             &  0.13 & \citet{s04} \\
             &  0.12 & \citet{sadakane05} \\
             &  0.11 & \citet{hm07} \\
             &  0.10 & Average \\
HD 139357    & -0.13 & \citet{d09b} \\
42 Dra       & -0.46 & \citet{d09b} \\    
HD 173416    & -0.22 & \citet{l09} \\
HD 180902    &  0.04 & \citet{j10} \\
HD 181342    &  0.26 & \citet{j10} \\
HD 240210    & -0.18 & \citet{n09b} \\
14 And       & -0.24 & \citet{sato08b} \\            
HD 222404    & +0.18 & \citet{f04} \\
             & +0.16 & \citet{s04} \\
             & +0.17 & Average \\

\enddata
\end{deluxetable}


\end{document}